\documentclass{article}
\usepackage[final]{neurips_2023}
\usepackage[utf8]{inputenc} 
\usepackage[T1]{fontenc}    
\usepackage{hyperref}       
\usepackage{url}            
\usepackage{booktabs}       
\usepackage{amsfonts}       
\usepackage{nicefrac}       
\usepackage{microtype}      
\usepackage{lipsum}
\usepackage{amsmath}
\usepackage{stackengine}
\usepackage{float}
\usepackage{amssymb}
\usepackage{graphicx}
\graphicspath{{Images/}}
\usepackage{subfig}
\usepackage{float}
\usepackage{geometry}
\usepackage{color, colortbl}
\usepackage{rotating}
\usepackage{subfig}
\usepackage{tikz}
\usepackage{xcolor}
\usepackage{animate}
\usepackage[sorting=none, style=nature]{biblatex} 
\bibliography{references} 

\title{Aberrant High-Order Dependencies in Schizophrenia Resting-State Functional MRI Networks}

\author{%
  Qiang Li$^{1*}$, Vince D. Calhoun$^{1}$, Adithya Ram Ballem$^{1}$, Armin Iraji$^{1,2*}$\\
  $^{1}$Tri-Institutional Center for Translational Research in Neuroimaging and Data Science (TReNDS)\\
  Georgia State University, Georgia Institute of Technology, Emory University, Atlanta, GA, USA \\
  $^{2}$Department of Computer Science, Georgia State University, Atlanta, GA, USA \\
  \And
  Shujian Yu$^{3}$\\
  $^{3}$Department of Computer Science \\ Vrije Universiteit Amsterdam, The Netherlands
  \And
  Jesus Malo$^{4}$\\
  $^{4}$Image Processing Laboratory\\
  University of Valencia, Spain
  }%

\begin{document}
\maketitle

\begin{abstract}
The human brain has a complex, intricate functional architecture. While many studies primarily emphasize pairwise interactions, delving into high-order associations is crucial for a comprehensive understanding of how functional brain networks intricately interact beyond simple pairwise connections. Analyzing high-order statistics allows us to explore the nuanced and complex relationships across the brain, unraveling the heterogeneity and uncovering patterns of multilevel overlap on the psychosis continuum. Here, we employed high-order independent component analysis (ICA) plus multivariate information-theoretical metrics ($O$-information and $S$-information) to estimate high-order interaction to examine schizophrenia using resting-state fMRI. The results show that multiple brain regions networks may be altered in schizophrenia, such as temporal, subcortical, and higher-cognitive brain regions, and meanwhile, it also shows that revealed synergy gives more information than redundancy in diagnosing schizophrenia. All in all, we showed that high-order dependencies were altered in schizophrenia. Identification of these aberrant patterns will give us a new window to diagnose schizophrenia.
\end{abstract}

\section{Introduction}   
Schizophrenia is a major psychotic disease that severely affects people's quality of life, and it is usually characterized by delusions, hallucinations, disorganization, and unusual behavior. The pathogenesis behind schizophrenia is complex, multifaceted, and involves multiple brain regions, and previous studies have shown that pair-wise functional connectomics is altered as one of the features of schizophrenia~\cite{FU2021117385,MENG2023103434,Du21commbio,Iraji23BioRxiv}. However, pair-wise functional connectomics will not capture multiple network statistical relationships and will ignore some high-order functional interactions that may play an important role in its use as a biomarker for schizophrenia~\cite{xie2021constructing,zhang2017hybrid,chen2016high}. The related studies have shown that multivariate mutual information can be used as a high-order functional connectivity descriptor and applied to the diagnosis of psychotic disorders~\cite{herzog2022genuine,lientropy21,li2022functional,li2022functionalen,li2022functionalnn,Licassp10193346,gatica2021high}. Therefore, in this study, we hypothesize that aberrant high-order dependencies are present in schizophrenia, and we investigate this using high model-order independent component analysis (ICA) plus information-theoretical metrics ($O$-information and $S$-information) to estimate high-order brain network interaction to examine schizophrenia using resting-state fMRI (rsfMRI).

\section{Materials and Methods}
\subsection{rsfMRI Dataset}
The total dataset of 1004 subjects from the Bipolar and Schizophrenia Network for Intermediate Phenotypes (BSNIP) consortium~\cite{tamminga2013clinical} (each half subject in typical controls and schizophrenia) used in this work was obtained by using 105 intrinsic connectivity network (ICN) (see \textbf{Appendices}, in Fig.~\ref{fig:3}) time courses derived from a multi-spatial-scale, spatially constrained ICA approach, and this allows generalizability and comparability of findings across studies. The shape of the time course is [T, 105], where T is the number of time points that are different between subjects~\cite{Iraji2022CanonicalAR}. The rsfMRI preprocessed steps as shown in Fig.~\ref{fig:1}: (a.) Quality control was applied to identify high-quality data. (b.) Each subject's rsfMRI data were preprocessed using a common procedure, including rigid body motion correction, slice timing correction, and distortion correction. (c.) Preprocessed subject data were registered into a common space, resampled to $3 mm^{3}$ isotropic voxels, and spatially smoothed using a Gaussian kernel with a $6 mm$ full width at half-maximum. (d.) We used a multi-spatial-scale template of 105 ICNs obtained from 100k+ subjects and a constraint ICA approach to obtain subject-specific ICN time courses, and they were cleaned using a common standard. (e.) The estimated 105 intrinsic connectivity networks are categorized into visual (VI), cerebellar (CB), temporal (TP), subcortical (SC), somatomotor (SM), and higher cognitive (HC) groups for the next information theory analysis.
\subsection{Information-Theoretical Analysis}

The \emph{Total Correlation (TC)} and \emph{Dual Total Correlation (DTC)} describe the dependence among $n$ variables and can be considered as a non-negative generalization of the concept of mutual information from two parties to $n$ parties. Let the definition of total correlation due Watanabe~\cite{watanabe1960information} be denoted as:
\begin{equation}\label{eq.tc}
\begin{split}
   & \mathbf{TC}\left(X^1, \cdots, X^{n}\right) = \sum_{i=1}^n \mathbf{H}\left(X^{i}\right)-\mathbf{H}\left(X^1, \cdots, X^{n}\right) \\
\end{split}
\end{equation}

Analogously with \emph{TC}, \emph{DTC} can be defined as~\cite{Han78},

\begin{equation}\label{eq.dtc}
\begin{split}
    & \mathbf{DTC}\left(X^1, \cdots, X^n\right)=\mathbf{H}\left(X^1, \cdots, X^n\right)-\sum_{i=1}^n \mathbf{H}\left(X^i \mid X^{[n] \backslash i} \right) \\
    &  =\left[\sum_{i=1}^n \mathbf{H}\left(X^{[n] \backslash i}\right)\right]-(n-1) \mathbf{H}\left(X^1, X^2, \cdots, X^n\right)
\end{split}
\end{equation}

where $X^{[n] \backslash i}= \{X^1, \cdots, X^{i-1}, X^{i+1}, \cdots, X^n\}$, i.e., the set of all variables excluding $X^i$. From these definitions in Eqs.~\ref{eq.tc},~\ref{eq.dtc}, if all variables are independent, both \emph{TC} and \emph{DTC} will be zero.

From the \emph{TC} and \emph{DTC}, we get $O$-Information and $S$-Information, which capture the overall balance between redundant and synergistic high-order dependencies in multivariate system~\cite{rosas2019quantifying}.

The $O$-Information of a set of $n$ random variables $X^{n}$ can be get as,

\begin{equation}
    \mathbf{\Omega_{n}}\left(X^1, \cdots, X^n\right)=\mathbf{T C}\left(X^1, \cdots, X^n\right)-\mathbf{D T C}\left(X^1, \cdots, X^n\right)
    \label{Eq.o}
\end{equation}

The $S$-Information of a set of $n$ random variables can be calculate as,  

\begin{equation}
    \mathbf{\Sigma_{n}}\left(X^1, \cdots, X^n\right)=\mathbf{T C}\left(X^1, \cdots, X^n\right)+\mathbf{D T C}\left(X^1, \cdots, X^n\right)
    \label{Eq.s}
\end{equation}

if $\emph{TC}>\emph{DTC}$ or $\mathbf{\Omega}>0$, the system corresponds to redundancy-dominated dependencies, and in the other way around, $\emph{DTC}>\emph{TC}$ or $\mathbf{\Omega}<0$, the system corresponds to synergy-dominated dependencies~\cite{rosas2019quantifying}.

Consider the fMRI signal as a Gaussian distribution, as shown in Fig.~\ref{fig:1}C, and for a univariate Gaussian random variable $X \sim \mathcal{N}(\mu, \sigma)$, the entropy (given in nats) will be $\mathbf{H}^{\mathcal{N}}(X)=\frac{\ln \left(2 \pi e \sigma^2\right)}{2}$, and for a multivariate Gaussian distribution, the joint entropy will be $\mathbf{H}^{\mathcal{N}}(\left(X^1, \cdots, X^n\right)=\frac{\ln \left[(2 \pi e)^n|\Sigma|\right]}{2}$, where $|\Sigma|$ refers to the determinant of the covariance matrix of $\left(X^1, \cdots, X^n\right)$.  The for the multivariate Gaussian case, the mutual information between $\left(X^1, \cdots, X^n\right)$ and $\left(Y^1, \cdots, Y^n\right)$ is given by, $\mathbf{I}((X^1, \cdots, X^n); (Y^1, \cdots, Y^n))=\frac{1}{2 \ln 2} \ln \left[\frac{\left|\Sigma_(X^1, \cdots, X^n)\right|\left|\Sigma_(Y^1, \cdots, Y^n)\right|}{\left|\Sigma_{(X^1, \cdots, X^n) (Y^1, \cdots, Y^n)}\right|}\right]$. Finally, the Gaussian estimator for total correlation can be:

\begin{figure}
    \centering
    \includegraphics[width=0.93\textwidth, height=14cm]{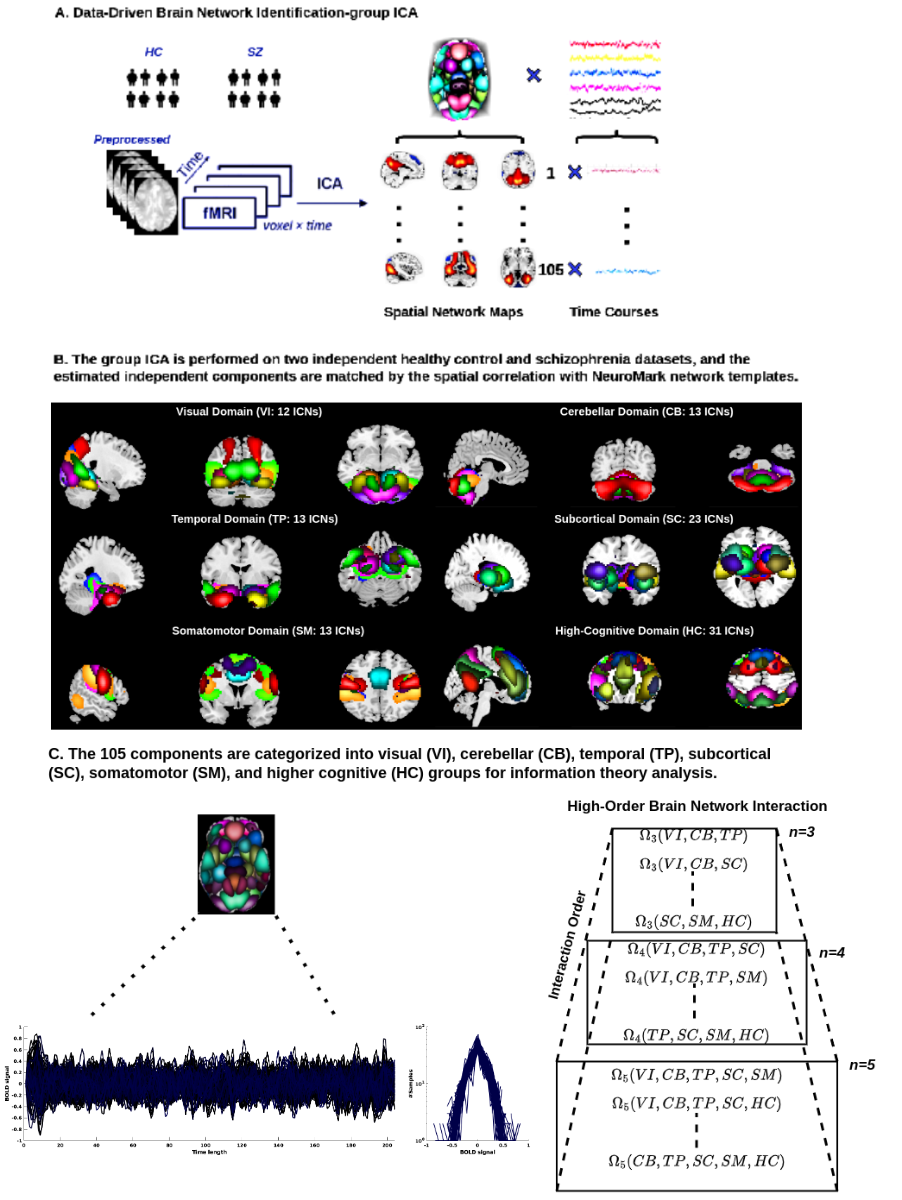}
    \caption{Flowchart of capturing high-order dependencies among brain connectivity states. \textbf{A.} Group ICA is performed on two independent healthy control and schizophrenia datasets, and corresponding 105 spatial network maps and time courses are extracted from each subject. \textbf{B.} The estimated independent components are matched with the NeuroMark\_2.1 template~\cite{Iraji2022CanonicalAR} based on spatial correlation, and then all components are categorized into six groups, i.e., VI, CB, TP, SC, SM, and HC. \textbf{C.} The BOLD signal from independent components matches a multivariate Gaussian distribution and related high-order (i.e., interaction order: 3, 4, 5) information measures with $O$-information and $S$-information.}
    \label{fig:1}
\end{figure}

\begin{equation}
    \mathbf{TC}^{\mathcal{N}}\left(X^1, \cdots, X^n\right)=\frac{-\ln (|\Sigma|)}{2}
    \label{Eq.gtc}
\end{equation}

From~\cite{Varley2022MultivariateIT}, the Eq.~\ref{Eq.o},~\ref{Eq.s} will be changed to,  

\begin{equation}
\mathbf{\Omega_{n}}^{\mathcal{N}}\left(X^1, \cdots, X^n\right)=(2-n) \mathbf{TC}^{\mathcal{N}}\left(X^1, \cdots, X^n\right)+\sum_{i=1}^n \mathbf{TC}^{\mathcal{N}}\left(X^{[n] \backslash i}\right)
\label{Eq.go}
\end{equation}
 
From Eq.~\ref{Eq.o},~\ref{Eq.s},~\ref{Eq.gtc},~\ref{Eq.go}, it can calculate all of the metrics described above, i.e., \emph{TC}, \emph{DTC}, \emph{O}-information and \emph{S}-information for multivariate Gaussian variables.

\section{Results and Discussion}
The low-order functional connectivity (LOFC) based on the pair-wise Pearson correlation in healthy control (HC) subjects and schizophrenia (SC) patients was present in Fig.~\ref{fig:2} and Fig.~\ref{fig:4} (see \textbf{Appendices}). We see that major aberrant brain network interaction happened between visual-somatomotor and temporal-high cognitive brain areas, and it reflects that three brain networks, such as visual, somatomotor, and higher cognitive, may be altered in the SC patients.

\begin{figure}[H]
    \centering
    \includegraphics[width=\textwidth, height=8.5cm]{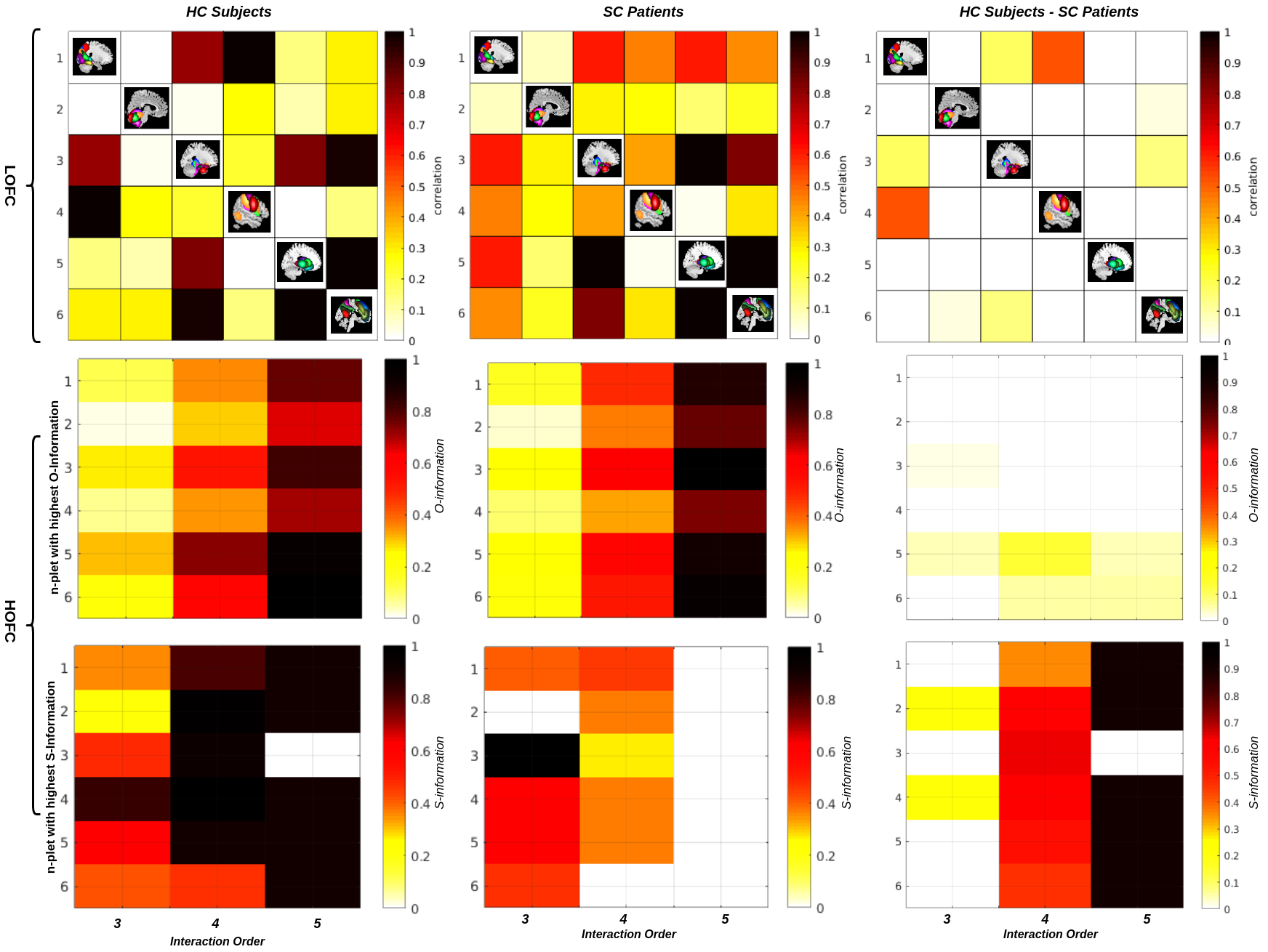}
    \caption{Comparison of low-order functional connectivity (LOFC) and high-order functional connectivity (HOFC) between HC subjects and SC patients. The top row shows that LOFC and the main difference present a pair-wise interaction between VI-SM and TP-HC. The bottom row presents the HOFC with redundancy and synergy information-theoretic measures made from HC subjects and SC patients, and the HOFC not only captures redundancy information but also synergy information in the human brain, and with interaction order increasing, we see that more difference is captured by the HOFC.}
    \label{fig:2}
\end{figure}

From the high-order functional connectivity (HOFC), the brain information networks integration and segregation both present from multivariate information theory measures (in Fig.~\ref{fig:2}), and we see that redundancy information basically mainly distributed at the SC and HC brain regions with interaction order gradually increases, and it also captures more aberrant brain networks, such as TP, SC, and HC brain regions, and it may suggest that these brain regions are involved in the SC patients. Moreover, synergy information was also uncovered by HOFC, and it showed that synergy information plays a more important role than redundancy in SC patients, and it may suggest that synergy is mainly aberrant in SC patients, as it can be a biomarker for diagnosing SC patients.

However, there are still some limitations, and the next step would be to dig into specific independent component networks and see which independent component networks mainly contribute to the SC patients. Second, explaining the role of redundancy and synergy in the SC patients and how to properly explain them from a neurophysics perspective.

\newpage
\printbibliography

@article{watanabe1960information,
  title={Information theoretical analysis of multivariate correlation},
  author={Watanabe, Satosi},
  journal={IBM Journal of research and development},
  volume={4},
  number={1},
  pages={66--82},
  year={1960},
  publisher={IBM}
}

@article{Han78,
  author = {Han, Te Sun},
  journal = {Inf. Control.},
  number = {2},
  pages = {133-156},
  title = {Nonnegative Entropy Measures of Multivariate Symmetric Correlations},
  volume = {36},
  year = {1978}
}

@article{li2022functional,
  title={Functional connectivity inference from fMRI data using multivariate information measures},
  author={Li, Qiang},
  journal={Neural Networks},
  volume={146},
  pages={85--97},
  year={2022},
  publisher={Elsevier}
}

@article{gatica2021high,
  title={High-order interdependencies in the aging brain},
  author={Gatica, Marilyn and Cofr{\'e}, Rodrigo and Mediano, Pedro AM and Rosas, Fernando E and Orio, Patricio and Diez, Ibai and Swinnen, Stephan P and Cortes, Jesus M},
  journal={Brain connectivity},
  volume={11},
  number={9},
  pages={734--744},
  year={2021},
  publisher={Mary Ann Liebert, Inc., publishers 140 Huguenot Street, 3rd Floor New~…}
}

@article{herzog2022genuine,
  title={Genuine high-order interactions in brain networks and neurodegeneration},
  author={Herzog, Rub{\'e}n and Rosas, Fernando E and Whelan, Robert and Fittipaldi, Sol and Santamaria-Garcia, Hernando and Cruzat, Josephine and Birba, Agustina and Moguilner, Sebastian and Tagliazucchi, Enzo and Prado, Pavel and others},
  journal={Neurobiology of Disease},
  volume={175},
  pages={105918},
  year={2022},
  publisher={Elsevier}
}

@article{li2022functionalen,
  title={Functional Connectome of the Human Brain with Total Correlation},
  author={Li, Qiang and Steeg, Greg Ver and Yu, Shujian and Malo, Jesus},
  journal={Entropy},
  volume={24},
  number={12},
  pages={1725},
  year={2022},
  publisher={Multidisciplinary Digital Publishing Institute}
}

@article{li2022functionalnn,
  title={Functional connectivity in visual areas from Total Correlation},
  author={Li, Qiang and Steeg, Greg Ver and Malo, Jesus},
  journal={arXiv preprint arXiv:2208.05770},
  year={2022}
}

@article{rosas2019quantifying,
  title={Quantifying high-order interdependencies via multivariate extensions of the mutual information},
  author={Rosas, Fernando E and Mediano, Pedro AM and Gastpar, Michael and Jensen, Henrik J},
  journal={Physical Review E},
  volume={100},
  number={3},
  pages={032305},
  year={2019},
  publisher={APS}
}

@article{FU2021117385,
title = {Dynamic state with covarying brain activity-connectivity: On the pathophysiology of schizophrenia},
journal = {NeuroImage},
volume = {224},
pages = {117385},
year = {2021},
author = {Zening Fu and Armin Iraji and Jessica A. Turner and Jing Sui and Robyn Miller and Godfrey D. Pearlson and Vince D. Calhoun}
}

@article{MENG2023103434,
title = {Multi-model order spatially constrained ICA reveals highly replicable group differences and consistent predictive results from resting data: A large N fMRI schizophrenia study},
journal = {NeuroImage: Clinical},
volume = {38},
pages = {103434},
year = {2023},
author = {Xing Meng and Armin Iraji and Zening Fu and Peter Kochunov and Aysenil Belger and Judy M. Ford and Sara McEwen and Daniel H. Mathalon and Bryon A. Mueller and Godfrey Pearlson and Steven G. Potkin and Adrian Preda and Jessica Turner and Theo G.M. {van Erp} and Jing Sui and Vince D. Calhoun}
}

@INPROCEEDINGS{Licassp10193346,
  author={Li, Qiang and Yu, Shujian and Madsen, Kristoffer H and Calhoun, Vince D and Iraji, Armin},
  booktitle={2023 IEEE International Conference on Acoustics, Speech, and Signal Processing Workshops (ICASSPW)}, 
  title={Higher-Order Organization in the Human Brain From Matrix-Based Rényi’s Entropy}, 
  year={2023},
  pages={1-5}
}

@inproceedings{lientropy21,
author = {Li, Qiang},
year = {2021},
month = {05},
pages = {9797},
booktitle={Entropy 2021: The Scientific Tool of the 21st Century},
title = {Measuring Functional Connectivity of Human Intra-Cortex Regions with Total Correlation}
}

@article{Du21commbio,
author = {Du, Yuhui and Fu, Zening and Xing, Ying and Lin, Dongdong and Pearlson, Godfrey and Kochunov, Peter and Qi, Shile and Salman, Mustafa and Abrol, Anees and Calhoun, Vince},
year = {2021},
month = {09},
pages = {},
title = {Evidence of shared and distinct functional and structural brain signatures in schizophrenia and autism spectrum disorder},
volume = {4},
journal = {Communications Biology}
}

@article{Iraji2022CanonicalAR,
  title={Canonical and Replicable Multi-Scale Intrinsic Connectivity Networks in 100k+ Resting-State fMRI Datasets},
  author={Armin Iraji and Zening Fu and Ashkan Faghiri and Marlena Duda and J. Chen and Srinivas Rachakonda and Thomas P DeRamus and Peter V. Kochunov and Bhim Mani Adhikari and Aysneil Belger and Judith M Ford and D. H. Mathalon and G. D. Pearlson and Steven G. Potkin and Adrian Preda and JA Turner and Theo G. M. van Erp and Juan R. Bustillo and K. Z. Yang and Koko Ishizuka and A. Sawa and Kent E. Hutchison and Elizabeth A. Osuch and Jean Th{\'e}berge and Christopher C. Abbott and Bryon A. Mueller and Dongmei Zhi and Chuan-jun Zhuo and S Liu and Yang Xu and Mustafa S. Salman and J. Liu and Y. Du and Jing Sui and T. Adalı and Vince D. Calhoun},
  journal={bioRxiv},
  year={2022},
  month = {09},
  doi = {10.1101/2022.09.03.506487}
}

@article{Varley2022MultivariateIT,
  title={Multivariate information theory uncovers synergistic subsystems of the human cerebral cortex},
  author={Thomas F. Varley and Maria Pope and Joshua Faskowitz and Olaf Sporns},
  journal={Communications Biology},
  year={2022},
  volume={6}
}

@article{Iraji23BioRxiv,
author = {Iraji, Armin and Chen, Jiayu and Lewis, Noah and Fu, Zening and Agcaoglu, Oktay and Kochunov, Peter and Adhikari, Bhim and Mathalon, Daniel and Pearlson, Godfrey and Macciardi, Fabio and Preda, A. and Erp, Theodorus and Bustillo, Juan and Díaz-Caneja, Covadonga and Andres-Camazon, Pablo and Dhamala, Mukesh and Adali, Tulay and Calhoun, Vince},
year = {2023},
month = {07},
title = {Spatial Dynamic Subspaces Encode Sex-Specific Schizophrenia Disruptions in Transient Network Overlap and its Links to Genetic Risk},
journal = {bioRxiv : the preprint server for biology}
}

@article{Coifman06acha,
author = {Coifman, Ronald and Lafon, Stéphane},
year = {2006},
month = {07},
pages = {5-30},
title = {Diffusion maps},
volume = {21},
journal = {Applied and Computational Harmonic Analysis}
}

@article{xie2021constructing,
  title={Constructing high-order functional connectivity network based on central moment features for diagnosis of autism spectrum disorder},
  author={Xie, Qingsong and Zhang, Xiangfei and Rekik, Islem and Chen, Xiaobo and Mao, Ning and Shen, Dinggang and Zhao, Feng},
  journal={PeerJ},
  volume={9},
  pages={e11692},
  year={2021},
  publisher={PeerJ Inc.}
}

@article{zhang2017hybrid,
  title={Hybrid high-order functional connectivity networks using resting-state functional MRI for mild cognitive impairment diagnosis},
  author={Zhang, Yu and Zhang, Han and Chen, Xiaobo and Lee, Seong-Whan and Shen, Dinggang},
  journal={Scientific reports},
  volume={7},
  number={1},
  pages={6530},
  year={2017},
  publisher={Nature Publishing Group UK London}
}

@article{chen2016high,
  title={High-order resting-state functional connectivity network for MCI classification},
  author={Chen, Xiaobo and Zhang, Han and Gao, Yue and Wee, Chong-Yaw and Li, Gang and Shen, Dinggang and Alzheimer's Disease Neuroimaging Initiative},
  journal={Human brain mapping},
  volume={37},
  number={9},
  pages={3282--3296},
  year={2016},
  publisher={Wiley Online Library}
}

@article{tamminga2013clinical,
  title={Clinical phenotypes of psychosis in the Bipolar-Schizophrenia Network on Intermediate Phenotypes (B-SNIP)},
  author={Tamminga, Carol A and Ivleva, Elena I and Keshavan, Matcheri S and Pearlson, Godfrey D and Clementz, Brett A and Witte, Bradley and Morris, David W and Bishop, Jeffrey and Thaker, Gunvant K and Sweeney, John A},
  journal={American Journal of psychiatry},
  volume={170},
  number={11},
  pages={1263--1274},
  year={2013},
  publisher={Am Psychiatric Assoc}
}

\newpage
\section{Appendices}
\subsection{Independent Component Networks}
\begin{figure}[H]
    \centering
    \includegraphics[width=\textwidth, height=18cm]{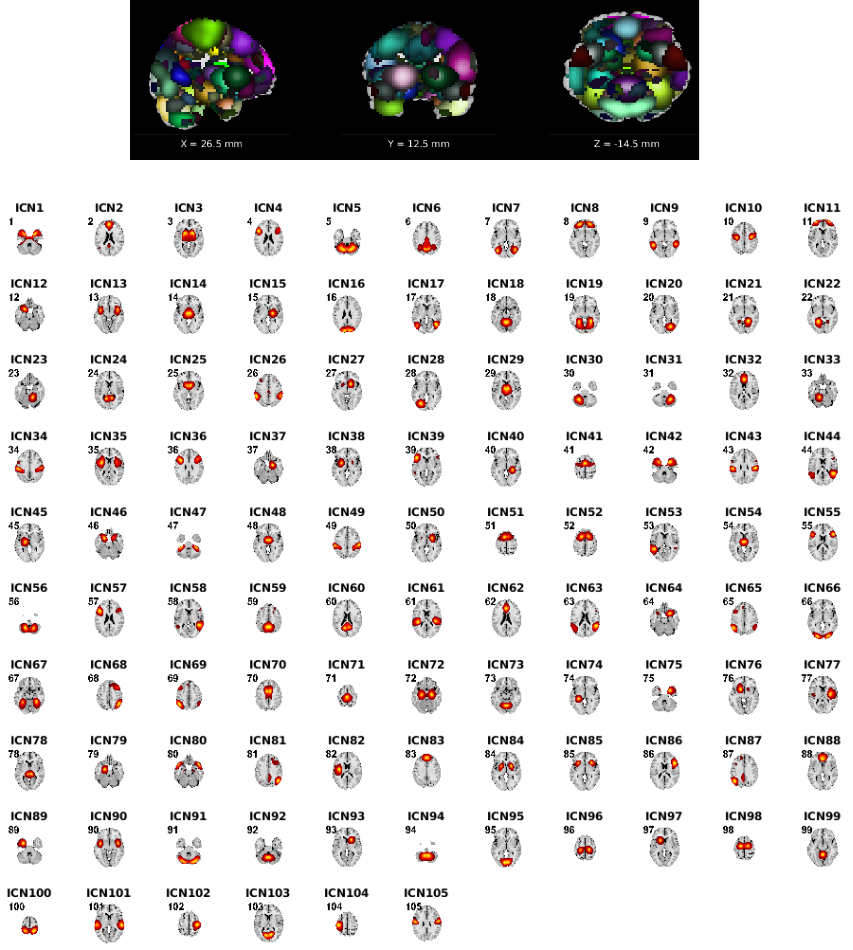}
    \caption{A total of 105 independent component networks are presented on the overlay brain volume, and different colors refer to each independent component network. Each component is also displayed separately.}
    \label{fig:3}
\end{figure}

Total of 105 independent component networks identified from multivariate spatially constrained ICA are shown in Fig.~\ref{fig:3}. 

\newpage

\subsection{Low-Order Functional Connectogram}
The low-order functional connectivity estimated for 105 independent component networks, and we applied the diffusion embedding algorithm~\cite{Coifman06acha} to cluster the related independent component networks.

\begin{figure}[H]
    \centering
    \includegraphics[width=\textwidth, height=11.6cm]{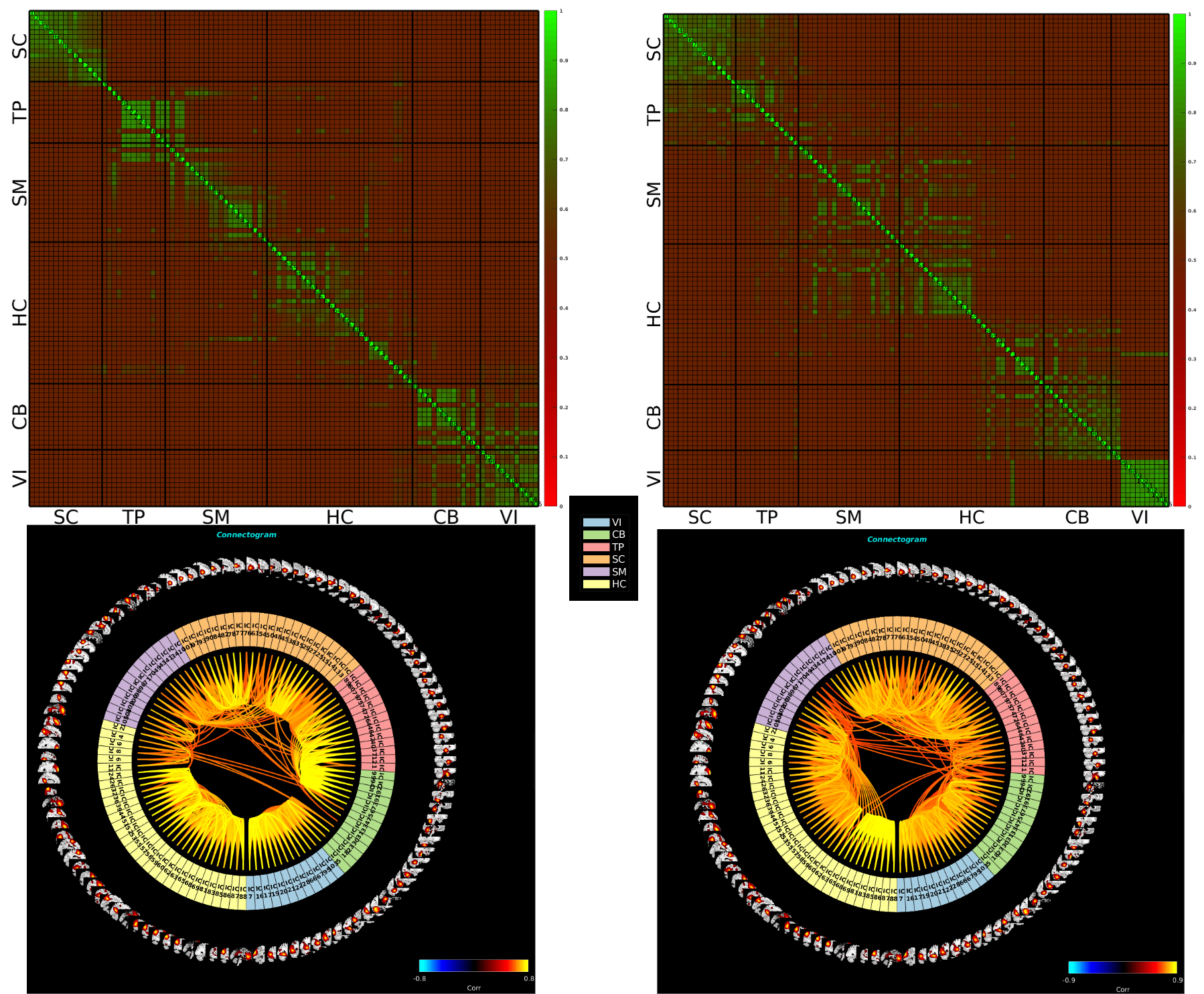}
    \caption{Comparison of clustering low-order functional connectivity (LOFC) between healthy controls (HC subjects, left side) and schizophrenia (SC patients, right side) The top row shows that LOFC presents a pair-wise interaction. The bottom row presents the functional connectogram corresponding to each correlation map with a threshold set to 0.6.}
    \label{fig:4}
\end{figure}

Here we clearly see that interaction-independent component network patterns are clumped, and the pattern can be matched to six multi model order ICA brain template networks (see Fig.~\ref{fig:4}). The corresponding functional connectogram was also presented with a threshold of 0.6.

\subsection{High-Order Functional Connectogram}
Given the number of independent component networks, exploring the high-order functional connectogram with 105 independent component networks along with increasing interaction order will face some challenges. Furthermore, it will also face how to present the result with a graph connectogram, and all these problems will be major problems that need to be solved, but it will be very important for us to precision diagnose schizophrenia, and the previous related research~\cite{li2022functionalen} has already provided us with a window to solve it, and we can apply it to the precision diagnosis of schizophrenia.

\end{document}